\documentstyle[twocolumn,aps]{revtex}
\def\ibb #1{\leavevmode\hbox{\kern.3em\vrule
     height 1.5ex depth -.1ex width .4pt\kern-.3em\rm#1}}
\def\Cx {{\ibb C}}
\def\Fx {{\ibb F}}
\def\QED{\leavevmode\unskip\penalty9999 \hbox{}\nobreak\hfill
     \quad\hbox{\leavevmode  \hbox to.77778em{%
               \hfil\vrule   \vbox to.675em%
               {\hrule width.6em\vfil\hrule}\vrule\hfil}}
     \par\vskip14pt}

\def\tr{\mathop{\rm tr\!}\nolimits}

\newtheorem{Theo}{Theorem}

\newtheorem{Lem}{Lemma}
\begin{document}
\title{Distillability via protocols respecting the positivity of partial transpose}
\author{Tilo Eggeling\thanks{Electronic Mail:
\tt{T.Eggeling@tu-bs.de}}, Karl Gerd H.
Vollbrecht\thanks{Electronic Mail: \tt{K.Vollbrecht@tu-bs.de}},
Reinhard F. Werner\thanks{Electronic Mail:
\tt{R.Werner@tu-bs.de}}, Michael M. Wolf\thanks{Electronic Mail:
\tt{MM.Wolf@tu-bs.de}}}
\address{
Institut f\"ur Mathematische Physik, TU Braunschweig\\
Mendelssohnstr.3, 38106 Braunschweig, Germany.}

\date{\today}

\maketitle

\begin{abstract}We show that all quantum states that do not have a
positive partial transpose are distillable via channels, which
preserve the positivity of the partial transpose. The question
whether NPT bound entanglement exist is therefore closely related
to the connection between the set of separable superoperators and
PPT-preserving maps.
\end{abstract}

\section{Introduction}

One of the main tasks of quantum information theory is the
systematic investigation of quantum entanglement, which is one of
the key ingredients in quantum computation and quantum information
processing. In spite of considerable research efforts, however,
there are still many aspects of entanglement which are not fully
understood. This is not only true for the quantitative theory (the
explicit computation or at least the estimation of entanglement
measures) but even for qualitative features.

These qualitative features are best explained by looking at the
history of the problem. In 1989 \cite{W89} it was a new
realization that there is a proper gap between obviously entangled
states (those violating some Bell inequality) and the obviously
non-entangled states, which are now called {\it separable}. The
next step, made by Sandu Popescu in 1995 \cite{P95} was the
striking result that this gap could be narrowed by {\it
distillation}: By local filtering and classical communication one
could sometimes get highly entangled states even from states not
violating any Bell inequality. For a while it was everybody's
favourite conjecture that there should be no more gap, i.e., that
all non-separable states should be distillable. This folk
conjecture was shattered in 1998 by counterexamples \cite{HHH2},
which are now called {\it bound entangled} states. The way these
examples were established was by showing that the property of a
density operator of having a positive partial transpose, i.e., of
being a {\it ppt-state}, does not change under distillation.
Therefore any non-separable state with positive partial transpose
has to be ``bound entangled''. The obvious white spot on the
entanglement map is then: Are these all bound entangled states, or
are there undistillable states, whose partial transpose is not
positive?

There were two recent papers \cite{DD,BL,problemsite} presenting
some evidence for the existence of non-ppt bound entangled states.
However, the matter is not decided \cite{Gaussians}, and in view
of the rapidly growing dimensions of the Hilbert spaces involved,
numerical evidence can be treacherous in this field. The latest
development was an attempt by Pawel Horodecki \cite{Pawel} at
showing the existence of a gap between ppt-states and distillable
states, using a stronger protocol \cite{Rains} of distillation.
The attempt failed due to an error in another paper, but it
remained unclear whether the idea could be made to work. What we
show in the present note is that it cannot work: using the same
distillation protocol \cite{Rains}, every non-ppt-state becomes
distillable.

The rather subtle dependence of distillability on ``protocols''
requires some explanation. Typically a protocol fixes the amount
of classical communication allowed to Alice and Bob in the
process. Thus we may distinguish distillation with no
communication allowed or with one-way or two-way communication.
Even stronger protocols than two-way communication protocols
exist: these are defined by requiring only a subset of the
properties which are true for all two-way distillation procedures.
One example is the requirement that the overall operation can be
written as a sum of tensor products of local operations
(``separable superoperator''). Another such property, which is the
one we consider in this paper following Rains \cite{Rains}, is
that operators with positive partial transpose are again taken to
such operators. An example of such a ppt-preserving protocol is
the case where Alice and Bob share a ppt bound entangled state and
use a protocol consisting of local operations and classical
communication (LOCC) \cite{terhal}.

Obviously, the weaker the requirements on the admissible
transformations, the larger the set of distillable states.
However, when we do not care about rates of distillation, the
dependence on the protocol is not as strong as one might think.
For example, distillability with two-way communication and with
separable superoperators are known to be equivalent \cite{HHH2}.
Moreover the stronger protocols have the virtue of being much more
manageable and more easily parametrized than two-way communication
processes, which may involve an arbitrarily large number of
exchanges of classical information.

Therefore it seemed quite reasonable to study the problem of a
proper gap between ppt states and distillable states under this
 ``ppt preserving'' protocol. Moreover, since a proper gap is the
currently favored conjecture, it was reasonable to expect a gap
even with such a protocol. The main result of this note is,
however, that the gap disappears, if we allow such a strong
protocol. Unfortunately, this does not provide conclusive evidence
about the gap for weaker protocols.

\section{Distillation via PPT-preserving channels}

For the sake of completeness we begin by recapitulating the result
of Rains \cite{Rains} for the {\it fidelity of distillation} via
PPT-preserving channels. Let $\rho$ be a density operator
corresponding to a quantum state on $\Cx^d\otimes\Cx^d$ and
$\rho\mapsto T(\rho)$ a trace-preserving positive map, such that
$\sigma\geq0$ implies $T(\sigma^{T_2})^{T_2}\geq 0$, where the
superscript $ ^{T_2}$ denotes the partial transposition with
respect to a given basis. Let us further write
$P_m=|\psi_m\rangle\langle\psi_m|$ for the projector onto the
maximally entangled state in $m\times m$ dimensions, i.e.
$|\psi_m\rangle=\frac1{\sqrt{m}}\sum_{i=1}^m|i\rangle\otimes|i\rangle$.
Theorem 3.1 of \cite{Rains} then reads:

\begin{Lem}\label{Lem1}
The maximal fidelity of distillation via PPT and trace preserving
positive maps with respect to the $m$-dimensional maximally
entangled state is given by
\begin{equation}\label{Rainsbound}
F_m(\rho):=\max_T{\rm tr}\big[P_mT(\rho)\big] = \max_A {\rm tr}\big[\rho
A\big] ,
\end{equation}
where the maximum on the right side is taken over all hermitian
operators $A$ satisfying
\begin{equation}
0\leq A\leq{\bf 1} \quad\mbox{and}\quad-{\bf 1}\leq m A^{T_2} \leq
{\bf 1}.
\end{equation}
\end{Lem}
{\it Proof : \ }First note that since every unitary of the form
$U\otimes\overline{U}$ commutes with $P_m$ it suffices to consider
trace preserving positive maps mapping into the set of isotropic
states, i.e., states, which are obtained by averaging over all
these unitaries \cite{WV}:
\begin{equation}\label{Trho}
T(\rho) = {\rm tr}[\rho B] \big({\bf 1}-P_m\big) + {\rm tr}[\rho
A] P_m.
\end{equation}
The coefficients in Eq.(\ref{Trho}) have to be linear functionals
of $\rho$ so that we are free to write them as traces, and
$T(\rho)$ being again a proper state requires that $0\leq A,B
\leq{\bf 1}$, and $(m^2-1)B+A={\bf 1}$. In order to obtain a
PPT-preserving channel we have additionally to demand that
$\sigma\geq 0$ implies that
\begin{equation}\label{PPTpreserving}
T(\sigma^{T_2})^{T_2} = {\rm tr}[\sigma B^{T_2}] \big({\bf
1}-\frac{1}{m}\Fx\big) + {\rm tr}[\sigma A^{T_2}] \frac{1}{m}\Fx
\geq 0,
\end{equation}
where $\Fx$ denotes the flip operator, i.e.,
$\Fx|\phi\rangle\otimes|\psi\rangle =
|\psi\rangle\otimes|\phi\rangle$. Inequality (\ref{PPTpreserving})
is satisfied iff the absolute value of the coefficient of the flip
operator is less or equal than the weight of the identity
operator:
\begin{equation}\label{flipvsidentity}
\pm \frac{1}{m} {\rm tr}\Big[\sigma\Big(\frac{A^{T_2}-{\bf
1}}{m^2-1}+A^{T_2}\Big)\Big] \leq {\rm tr}\Big[ \sigma\frac{{\bf
1}-A^{T_2}}{m^2-1}\Big].
\end{equation}
Since this inequality has to hold for all positive operators
$\sigma$  we can reformulate it as an operator inequality which is
in turn equivalent to $-{\bf 1}\leq m A^{T_2} \leq {\bf 1}$.

Hence, there is a one-to-one correspondence between PPT-preserving
maps $T$ of the form (\ref{Trho}) and the respective hermitian
operators $A$ satisfying the constraints specified in Lemma
\ref{Lem1} given by ${\rm tr}\big[P_mT(\rho)\big] =  {\rm
tr}\big[\rho A\big] $.\QED

In fact positive maps of the form
(\ref{Trho}) are even completely positive, i.e. Lemma \ref{Lem1}
holds also for ppt-pre\-ser\-ving channels as can easily be seen by writing down a Kraus
decomposition:
\begin{eqnarray*}
T(\rho) &=& {\rm tr}[\rho B] \big({\bf 1}-P_m\big) + {\rm tr}[\rho A] P_m\\
&=&\left({\bf 1}-P_m\right)\tr{\left[\sqrt{B}\rho\sqrt{B}\right]}\left({\bf 1}-P_m\right)\\
&&+ P_m\tr{\left[\sqrt{A}\rho\sqrt{A}\right]}P_m.
\end{eqnarray*}
This however is a special property of positive maps of the form
(\ref{Trho}). Indeed positivity and ppt-preservation do not imply
complete positivity in general, a counterexample being the
transposition.

Now we can utilize Lemma \ref{Lem1} in order to prove the following
\begin{Theo} Any NPT state, i.e., state with not positive partial
transpose, is distillable via PPT-preserving channels.
\end{Theo}
{\it Proof : \ }We recall that a state is known to be distillable
via standard LOCC distillation protocols if ${\rm tr}[\rho
P_m]>\frac1m$ \cite{oneoverd}. The task is therefore to find an
appropriate operator $A$ such that ${\rm tr}[\rho A]>\frac1m$.

Let $P_{neg}$ be the projector onto the negative eigenspace of
$\rho^{T_2}$. We choose $A$ to be of the form
\begin{equation}\label{A}
A=\frac1m\big({\bf 1} - \epsilon P_{neg}^{T_2}\big), \quad
0<\epsilon\leq \min\Big\{2,||P_{neg}^{T_2}||_\infty^{-1}\Big\},
\end{equation}
where $||\cdot||_\infty$ denotes the operator norm. Now we have to
check, whether $A$ satisfies the constraints in Lemma \ref{Lem1}.

Positivity of the parameter $\epsilon$ implies $mA^{T_2} \leq {\bf
1}$. To ensure $A \leq {\bf 1}$ it is sufficient that $\epsilon
\leq (m-1)||P_{neg}^{T_2}||_\infty^{-1}$ but $0 \leq A $ requires
the even stronger condition
$\epsilon\leq||P_{neg}^{T_2}||_\infty^{-1}$. Moreover, $mA^{T_2}
\geq -{\bf 1}$ is equivalent to $\epsilon\leq 2$, which shows that
Eq.(\ref{A}) indeed defines an admissible operator $A$. With the
above $A$ we obtain
\begin{equation}\label{trrhoA}
{\rm tr}\big[\rho A\big]=\frac{1 + \epsilon{\cal N}(\rho)}m,
\end{equation}
where ${\cal N}(\rho)$ is the {\it negativity} \cite{negativity},
which is just the sum over the absolute values of the negative
eigenvalues of $\rho^{T_2}$. Since the state has at least one such
negative eigenvalue by assumption, we end up with a fidelity
larger than $\frac1m$, which completes our proof. \QED

Of course one may further evaluate Eq. (\ref{trrhoA}) for more
specific states. Let us for instance consider states commuting
with all unitaries of the form $U\otimes U$ \cite{W89}, which can
be written as
\begin{equation}\label{Werner}
\rho(p)= (1-p)\frac{P_+}{r_+} + p \frac{P_-}{r_-} , \quad 0\leq p
\leq 1,
\end{equation}
where $P_+$ ($P_-$) is the projector onto the symmetric
(antisymmetric) subspace of $\Cx^d\otimes\Cx^d$ and $r_\pm = {\rm
tr}[P_\pm] = \frac{d^2\pm d}2$ are the respective dimensions.
Evaluating Eq.(\ref{trrhoA}) for these states ($\epsilon=2$) then
leads to
\begin{equation}\label{Wernerfidelity}
{\rm tr}[A\rho(p)]=\frac{||\rho(p)^{T_2}||_1}m =
\frac{d-2+4p}{md}.
\end{equation}
In fact this turns out to be already the maximal value for
$F_m(\rho(p))$. This can easily be seen by decomposing the partial
transpose of the state into its positive and negative part, i.e.,
$\rho^{T_2}=\rho_+-\rho_-$. Then
\begin{eqnarray}\label{optbound}
{\rm tr}\big(\rho A\big)&=&{\rm tr}\big(A^{T_2}\rho^{T_2}
\big)={\rm tr}\big(A^{T_2}(\rho_+-\rho_-)\big)\nonumber\\  &\leq&
\frac1m {\rm tr}(\rho_++\rho_-)=\frac1m ||\rho^{T_2}||_1 ,
\end{eqnarray}
where the estimate is due to the constraint $m ||A||_1\leq 1$.

In fact this bound for the maximal fidelity can always be reached
for states with $ ||P_{neg}^{T_2}||_\infty \leq \frac12$.

\section{Conclusion}
We have argued that enlarging the set of distillation protocols to
PPT-preserving channels immediately implies that any NPT state can
be distilled. Since we know that a state can be distilled via
proper LOCC operations iff ${\rm tr}[P_m S(\rho)]>\frac1m$ for
some {\it separable superoperator} $S$ \cite{HHH2,RainsSSO}, this
raises the question about the connection between the sets of
separable superoperators and PPT-preserving channels. It is
obvious that any separable superoperator is PPT-preserving but
we do not know yet any efficient method for deciding whether a given
operator $A$ from Lemma \ref{Lem1} corresponds to a separable
superoperator.

There is a standard argument telling us that   NPT bound entangled
states exist iff there exist undistillable entangled states of the
form special $U\otimes U$-invariant form (\ref{Werner})
\cite{oneoverd}. So the question about the existence of {\it NPT
bound entangled states} becomes to decide whether PPT-preserving
channels that distill $U\otimes U$-invariant states near the
separable boundary can be realized as separable superoperators or
not.

Moreover, the above discussion raises the question whether it
suffices to use LOCC operations and PPT bound entangled states as
an additional resource in order to distill all NPT states.

Another interesting feature of the distillation we discussed is
that we only needed a {\it single copy} of the given bipartite
state, and not a tensor product of many identically prepared ones.
This raises the question whether distillability under LOCC
protocols can also be decided at the single copy level. All
examples known to us would be consistent with this.

\section*{Acknowledgement}
The authors would like to thank Barbara Terhal for interesting
discussions and for pointing out reference \cite{terhal}.
Funding by the European Union project EQUIP (contract
IST-1999-11053) and financial support from the DFG (Bonn) is
gratefully acknowledged.


\begin{references}
\bibitem{W89} R.F. Werner, Phys. Rev. A, {\bf 40}, 4277 (1989).
\bibitem{P95} S. Popescu, Phys. Rev. Lett. {\bf 74}, 2619 (1995).
\bibitem{HHH2} M. Horodecki, P. Horodecki, and R. Horodecki, Phys. Rev. Lett. {\bf
 80}, 5239 (1998).
\bibitem{DD}  D.P. DiVincenzo, P.W. Shor, J.A. Smolin, B.M. Terhal, and A.V.
     Thapliyal,  Phys. Rev. A {\bf 61}, 062312 (2000).
\bibitem{BL}  W. D\"ur, J.I. Cirac, M. Lewenstein, D. Bruss, Phys. Rev. A {\bf 61}, 062313
(2000).
\bibitem{problemsite} See also problem page 2 at our open problem site {\tt
http://www.imaph.tu-bs.de/qi/problems/2.html}
\bibitem{Gaussians} An indication to the contrary is the very
recent result \cite{G2} that the gap between ppt and
distillability does not exist for Gaussian (continuous variable)
states.
\bibitem{G2} G. Giedke, L.-M. Duan, P. Zoller, and J.I. Cirac, quant-ph/0104072 (2001).
\bibitem{Pawel} P. Horodecki, quant-ph/0103091v1
(2001).
\bibitem{Rains} E.M. Rains, quant-ph/0008047 (2000).
\bibitem{terhal} P.W. Shor, J.A. Smolin and B.M. Terhal, Phys.
Rev. Lett. {\bf 86}, 2681 (2001).
\bibitem{WV} K.G.H. Vollbrecht and R.F. Werner, quant-ph/0010095
(2000).
\bibitem{oneoverd} M. Horodecki and P. Horodecki, Phys. Rev. A
{\bf 59}, 4206 (1999); S.L. Braunstein, C.M. Caves, R. Jozsa, N.
Linden, S. Popescu, and R. Schack, Phys. Rev. Lett. {\bf 83},
1054 (1999).
\bibitem{negativity} G. Vidal and R.F. Werner, quant-ph/0102117 (2001).
\bibitem{RainsSSO} E. Rains, quant-ph/9707002 (1997).

%\bibitem{EPR} A. Einstein, B. Podolsky, and N. Rosen, Phys. Rev. {\bf 47}, 777 (1935).
%\bibitem{Peres} A. Peres, Phys. Rev. Lett. {\bf 77}, 1413 (1996).
%\bibitem{HHH} M. Horodecki, P. Horodecki, and R. Horodecki, Phys. Lett. A {\bf 223}, 1 (1996).
%\bibitem{Lew} P. Horodecki, M. Lewenstein, G. Vidal, and I. Cirac, Phys. Rev. A {\bf 62}, 032310 (2000).
%\bibitem{Simon} R. Simon, Phys. Rev. Lett. {\bf 84}, 2726 (2000).
%\bibitem{WWGauss} R.F. Werner and M.M. Wolf, Phys. Rev. Lett {\bf 86}; quant-ph/0009118 (2000).
\end{references}
\end{document}